\documentclass[letter]{ptptex}

\title{
Comments on the radial plane waves%
}

\author{
Seiji \textsc{Sakoda}%
}

\inst{
Department of Applied Physics, National Defense Academy\\
Hashirimizu, Yokosuka city, Kanagawa 239-8686, Japan\\
sakoda@nda.ac.jp
}

\abst{
The orthogonality of the radial plane waves, introduced by
Fujikawa, turns out to be broken for the case of infinite volume. We
will find, though they become overcomplete, the concept of the radial
plane waves remains useful for constructing radial path integrals.
}

\PTPindex{060, 064}  

\begin{document}

\maketitle
\section{Introduction}
Formally Hermitian radial momentum operator and its
eigenfunctions(radial plane waves) were shown by Fujikawa\cite{KF08} to
be useful in constructing radial path integrals. In this letter 
we argue the orthogonality of such eigenfunctions in the limit of
infinite volume. 

\section{Orthogonality and completeness of radial plane waves}
Normalization of plane waves in a box($a\le x\le b$, $L=b-a$)
\begin{equation}
 \label{eq:norm01}
\frac{1}{L}\int_{a}^{b}\!\!dx\,e^{-2\pi i(m-n)x/L}=\delta_{m,n},\
m,n=0,\,\pm1,\,\pm2,\,\dots,
\end{equation}
is often utilized to define the normalization of them in the infinite 
volume as a limit $L\to\infty$; for $p=2\pi m/L$ and $p{'}=2\pi
n/L$, we define
\begin{equation}
 \label{eq:norm02}
\lim\limits_{L\to\infty}L\delta_{p,p{'}}=2\pi\delta(p-p{'})
\end{equation}
by making use of
\begin{equation}
 \label{eq:norm03}
\lim\limits_{L\to\infty}
\int_{-L/2}^{L/2}\!\!dx\,e^{-i(p-p{'})x}=2\pi\delta(p-p{'}).
\end{equation}
In this procedure, it is crucial we are able to shift $x$ to $x{'}=x-c$,
with $c=(b+a)/2$ being independent of $L$, in Eq.\eqref{eq:norm01}
before taking the limit $L\to\infty$.

The similar procedure for radial plane waves fails to yield the delta
function. We may make a change of variables from
$r$ to $r{'}=r-c{'}$($c{'}=R/2$) in
\begin{equation}
 \label{eq:norm04}
\frac{1}{R}\int_{0}^{R}\!\!dr\,
e^{-2\pi i(m-n)r/R}=\delta_{m,n},
m,n=0,\,\pm1,\,\pm2,\,\dots,
\end{equation}
to find
\begin{equation}
 \label{eq:norm04b}
\frac{1}{R}e^{-i(p-p{'})c{'}}\int_{-R/2}^{R/2}\!\!dr{'}\,
e^{-i(p-p{'})r{'}}=\delta_{p,p{'}},\
p=\frac{2\pi m}{R},\,p{'}=\frac{2\pi n}{R}.
\end{equation}
However, we cannot take the limit $R\to\infty$ in $R\delta_{p,p{'}}$
naively because $c{'}=R/2$ also tends to infinity for this case.
By treating the factor $e^{-i(p-p{'})R/2}$ with due care, we obtain
\begin{equation}
 \label{eq:norm04c}
\lim\limits_{R\to\infty}R\delta_{p,p{'}}=\pi\delta(p-p{'})
-i\mathrm{P}\frac{1}{p-p{'}}=
\frac{-i}{p-p{'}-i0_{+}},
\end{equation}
instead. Here $0_{+}$ designates a positive infinitesimal and the
principal value is denoted by the symbol $\mathrm{P}$.
Clearly, it is in accord with the definition of an integral 
\begin{equation}
 \label{eq:norm05}
\int_{0}^{\infty}\!\!r^{d-1}\,dr\,\frac{1}{2\pi r^{d-1}}e^{-i(p-p{'})r}=
\frac{1}{2\pi i}\frac{1}{p-p{'}-i0_{+}}
\end{equation}
by means of the regularization $p\to p+i0_{+}$ in $e^{ipr}$.
Therefore the foregoing shows that the formally Hermitian
operator\cite{Arthurs}
\begin{equation}
 \label{eq:pr}
\hat{p}_{r}=-i\frac{1}{r^{(d-1)/2}}\frac{d}{dr}r^{(d-1)/2}
\end{equation}
requires a positive infinitesimal in the imaginary part of its
eigenvalues when considered on the whole positive half line
$0<r<\infty$.
Although they are no longer orthogonal, completeness of
radial plane waves still holds:
\begin{equation}
 \label{eq:complete}
\frac{1}{2\pi(rr{'})^{(d-1)/2}}
\int_{-\infty}^{\infty}\!\!dp\,
e^{ip(r-r{'})}
=\frac{1}{(rr{'})^{(d-1)/2}}
\delta(r-r{'}).
\end{equation} 

Note here that we could have had a delta function instead of
Eq.\eqref{eq:norm04c} if we were allowed to extend the domain of $r$
into negative region. This explains why the radial plane waves become
overcomplete; there exists no room, though it is admissible in
classical Lagrangian, for changing $r$ to $-r$ in quantum mechanics.

\section{Summary}
We have clarified that radial plane waves cannot be orthonormal in the
limit of infinite volume.
In view of the use of radial plane waves in Ref.~\citen{KF08}, the
existence of the completeness Eq.\eqref{eq:complete} above seems to be
essential for the construction of radial path integrals.
Therefore losing the orthogonality will not be so
serious in this regard. On the contrary,
this overcomplete set is quite useful to find radial path
integrals in the conventional Lagrangian form with the extra quantum
potential\cite{EdwardsGulyaev} in a very simple and quick manner. 
This will be evident when we compare Fujikawa's method with 
the use of asymptotic forms of modified Bessel
functions in Ref.~\citen{EdwardsGulyaev}. 

\section*{Acknowledgements}
The author would like to thank Professor K.~Fujikawa for bringing the
topic to his attention. He is also grateful to Dr.~I.~Tsutsui for valuable
comments.

%


\begin{thebibliography}{99}
  
\bibitem{KF08}
K.~Fujikawa, \PTP{120,2008,181}.

\bibitem{EdwardsGulyaev}
S.~F.~Edwards and Y.~V.~Gulyaev, \JL{Proc. Roy. Soc.,A279,1964,229}.

\bibitem{Arthurs}
A.~M.~Arthurs, \JL{Proc. Roy. Soc., A313,1969,445}.
\end{thebibliography}
\end{document}